\documentclass[aps,prd,twocolumn,showpacs,amsmath,amssymb,nofootinbib,eqsecnum, preprintnumbers]{revtex4-1}

\newcommand{\tht}{\vartheta}
\newcommand{\eps}{\varepsilon}
\newcommand{\one}{\textsc{i}}

 \newcommand{\dm}{n}

\newcommand{\be}{\begin{equation}}
\newcommand{\ee}{\end{equation}}
\newcommand{\ba}{\begin{eqnarray}}
\newcommand{\ea}{\end{eqnarray}}

\newcommand{\beq}{\begin{equation}}
\newcommand{\eeq}{\end{equation}}
\newcommand{\beqa}{\begin{eqnarray}}
\newcommand{\eeqa}{\end{eqnarray}}
\newcommand{\nn}{\nonumber}

\newcommand{\hook}{\raisebox{-0.35ex}{\makebox[0.6em][r]
{\scriptsize $-$}}\hspace{-0.15em}\raisebox{0.25ex}{\makebox[0.4em][l]{\tiny
 $|$}}}

\newcommand{\even}{{\mathrm{e}}}
\newcommand{\odd}{{\mathrm{o}}}

\newcommand{\fiber}{\boldsymbol}

\newcommand{\lied}{\mathcal{L}}
\newcommand{\cd}[1]{\frac{\eth}{\partial{#1}}}
\newcommand{\cv}[1]{{\partial}_{#1}}

\newcommand{\Ric}{{\mathrm{Ric}}}

\newcommand{\lst}[1]{\langle{#1}\rangle}

\newcommand{\ef}{E}

\newcommand{\KV}[1]{\xi^{(#1)}}
\newcommand{\KT}[1]{k^{(#1)}}
\newcommand{\PKV}{\xi}
\newcommand{\CCKY}[1]{h^{(#1)}}

\newcommand{\A}[1]{A^{\!(#1)}}

\newcommand{\Bo}[1]{\mathcal{B}^{(#1)}}
\newcommand{\Vo}{\mathcal{V}}

\newcommand{\psc}[1]{\Psi_{#1}}
\newcommand{\chc}[1]{\mathcal{X}_{#1}}
\newcommand{\psf}[1]{\tilde\Psi_{#1}}
\newcommand{\chf}[1]{\tilde{\mathcal{X}}_{#1}}

\newcommand{\clfm}[1]{{\slash\mspace{-9.5mu}{#1}}}

\begin{document}
\title{Dirac Equation in Kerr-NUT-(A)dS Spacetimes: Intrinsic Characterization of Separability in All Dimensions}
\author{Marco Cariglia}
\email{marco@iceb.ufop.br}
\affiliation{Universidade Federal de Ouro Preto, ICEB, Departamento de F\'isica.
  Campus Morro do Cruzeiro, Morro do Cruzeiro, 35400-000 - Ouro Preto, MG - Brasil}

\author{Pavel Krtou\v{s}}
\email{Pavel.Krtous@utf.mff.cuni.cz}
\affiliation{Institute of Theoretical Physics,
Faculty of Mathematics and Physics, Charles University in Prague,
V~Hole\v{s}ovi\v{c}k\'ach 2, Prague, Czech Republic}

\author{David Kubiz\v n\'ak}
\email{D.Kubiznak@damtp.cam.ac.uk}
\affiliation{DAMTP, University of Cambridge, Wilberforce Road, Cambridge CB3 0WA, UK}

\date{April 19, 2011}  

\begin{abstract}
We intrinsically characterize separability of the Dirac equation in Kerr-NUT-(A)dS spacetimes in all dimensions. Namely, we explicitly demonstrate that in such spacetimes there exists a complete set of first-order mutually commuting operators, one of which is the Dirac operator, that allows for common eigenfunctions which can be found in a separated form and correspond precisely to the general solution of the Dirac equation found by Oota and Yasui [arXiv:0711.0078]. Since all the operators in the set can be generated from the principal conformal Killing--Yano tensor, this establishes the (up to now) missing link among the existence of hidden symmetry, presence of a complete set of commuting operators, and separability of the Dirac equation in these spacetimes.
\end{abstract}

\preprint{DAMTP-2011-23}

\maketitle

\section{Introduction}\label{sc:intro}
The most general known stationary higher-dimensional vacuum
(including a cosmological constant) black hole
spacetimes with spherical horizon topology \cite{ChenEtal:2006cqg} possess many remarkable properties, some of which are directly
inherited from the four-dimensional Kerr-NUT-(A)dS geometry \cite{Carter:1968cmp}. The similarity stems from the existence of a hidden symmetry associated with the {\em principal conformal Killing--Yano (PCKY) tensor} \cite{KubiznakFrolov:2007}.
Such a symmetry generates the whole tower of explicit and hidden symmetries \cite{KrtousEtal:2007jhep} which in their turn are responsible for many of the properties,
including complete integrability of the geodesic motion \cite{PageEtal:2007, KrtousEtal:2007prd, HouriEtal:2008a}, special algebraic type of the Weyl tensor
\cite{ColeyEtal:2004a, HamamotoEtal:2007}, the existence of a Kerr--Schild form \cite{ChenLu:2008}, and separability of various field perturbations.
For reviews on the subject we refer to \cite{FrolovKubiznak:2008, YasuiHouri:2011}.

Especially interesting is a relationship between the existence of the PCKY tensor and separability of test field equations in the background of general Kerr-NUT-(A)dS spacetimes \cite{ChenEtal:2006cqg}. Namely,
explicit separation of the Hamilton--Jacobi and Klein--Gordon equations was demonstrated by Frolov {\em et al}  \cite{FrolovEtal:2007}
and the achieved separability was {\em intrinsically characterized} by  Sergyeyev and Krtou{\v s} \cite{SergyeyevKrtous:2008}.
In their paper, the latter authors demonstrated that in Kerr-NUT-(A)dS spacetimes in all dimensions there exists a  complete set of first-order and second-order
operators (one of which is the Klein--Gordon operator) that mutually commute.
These operators are constructed from 2nd-rank Killing tensors and Killing vectors that can all be generated from the PCKY tensor.
The common eigenfunction of these operators is characterized by operators' eigenvalues and can be found in a separated form---it
is precisely the separated solution obtained by Frolov {\em et al}.
In fact, the demonstrated results provide a textbook example of general theory discussed in \cite{BenentiFrancaviglia:1979, KalninsMiller:1981, DuvalValent:2005}.

Higher spin perturbations of Kerr-NUT-(A)dS spacetimes were also studied. Namely, general separation of the Dirac equation in all dimensions was demonstrated by Oota and Yasui \cite{OotaYasui:2008}, electromagnetic perturbations in $n=5$ spacetime dimensions were studied  in \cite{MurataSoda:2008},
separability of the linearized gravitational perturbations was with increasing generality studied in
\cite{KodamaIshibashi:2003, KunduriEtal:2006, MurataSoda:2008, OotaYasui:2010, Dias:2010a}.
(The study of gravitational perturbations is very important for example for establishing the (in)stability of higher-dimensional black holes, see, e.g., recent papers \cite{Dias:2010a, DiasEtal:2010} and references therein.)
We also mention a recent paper \cite{DurkeeReall:2011} on general perturbation theory in higher-dimensional algebraically special spacetimes which employs the higher-dimensional GHP formalism \cite{DurkeeEtal:2010} and attempts to generalize Teukolsky's results \cite{Teukolsky:1972, Teukolsky:1973}.

The aim of the present paper is to intrinsically characterize the result of Oota and Yasui \cite{OotaYasui:2008}. Namely, we want to demonstrate
that, similar to the Klein--Gordon case \cite{SergyeyevKrtous:2008}, separability of the Dirac equation in Kerr-NUT-(A)dS spacetimes in all dimensions
is underlaid by the existence of a complete set of mutually commuting operators,  one of which is the Dirac operator.
The corresponding set of operators was already studied and the mutual commutation proved in \cite{CarigliaEtal:2011};
the operators are of the first-order and correspond to Killing vectors and closed conformal Killing--Yano tensors---all generated
from the PCKY tensor. In this paper we pick up the threads of these results and demonstrate that, in a properly chosen representation,
the common eigenfunction of the symmetry operators in the set can be chosen in the tensorial R-separated form and corresponds precisely to the separated solution
of the Dirac equation found by Oota and Yasui  \cite{OotaYasui:2008}. Our paper generalizes the $n=4$ results of Chandrasekhar \cite{Chandrasekhar:1976} and Carter and McLenghan \cite{CarterMcLenaghan:1979} and
$n=5$ results of Wu \cite{Wu:2008, Wu:2008b}.

A plan of the paper is as follows.
In Sec.~\ref{sec2} we review the theory of the Dirac equation in curved spacetime while concentrating on the commuting symmetry operators
of the Dirac operator.
In Sec.~\ref{sec3} we introduce the Kerr-NUT-(A)dS spacetimes in all dimensions and summarize their basic properties. Sec.~\ref{sc:csop} is devoted to the discussion
of a complete set of the Dirac symmetry operators; an explicit representation of these operators is found. Sec.~\ref{sc:sep} is the principal section of the
paper where the tensorial R-separability of the Dirac
equation is discussed and the main assertion of the paper is proved. In Sec.~\ref{sc:sep2} we comment on the possibility of introducing a different representation of $\gamma$ matrices
in which the standard tensorial separability occurs. Sec.~\ref{concl} is devoted to conclusions. In App.~\ref{apx} we gather necessary technical results.

\section{Dirac equation in curved space}\label{sec2}
\subsection{Dirac bundle}
In what follows we write the dimension of spacetime as
\be
\dm = 2N + \eps\,,
\ee
with $\eps =0, 1$ parameterising the even, odd dimension, respectively. The Dirac bundle ${\fiber{D}M}$ has fiber dimension ${2^N}$. If necessary, we use capital Latin indices for tensors from the Dirac bundle. It is connected with the tangent bundle ${\fiber{T}M}$ of the spacetime manifold ${M}$ through the abstract gamma matrices ${\gamma^a\in \fiber{T}M\otimes\fiber{D}^1_1 M}$,
which satisfy
\begin{equation}\label{gamgamg}
    \gamma^a\gamma^b+\gamma^b\gamma^a=2\,g^{ab}\;.
\end{equation}
They generate an irreducible representation of the abstract Clifford algebra on the Dirac bundle. All linear combinations of products of the abstract gamma matrices (with spacetime indices contracted) form the Clifford bundle ${\fiber{Cl}M}$, which is thus identified with the space ${\fiber{D}^1_1M}$ of all linear operators on the Dirac bundle. The Clifford multiplication (`matrix multiplication') is denoted by juxtaposition of the Clifford objects. The gamma matrices also provide the Clifford map ${\gamma_*}$, the isomorphism of the exterior algebra ${\fiber{\Lambda}M}$ and of the Clifford bundle,
\begin{equation}\label{cleaiso}
    \clfm\omega \equiv \gamma_* \omega
      \equiv \sum_p \frac1{p!}\, (\omega_p)_{a_1\dots a_p}\gamma^{a_1\dots a_p}\;.
\end{equation}
Here, ${\omega = \sum_p  \omega_p\in\fiber{\Lambda}M}$ is an inhomogeneous form, $\omega_p$ its $p$-form parts, $\omega_p\in \fiber{\Lambda}^p M$, and ${\gamma^{a_1\dots a_p} = \gamma^{[a_1}\cdots\gamma^{a_p]}}$.
For future use we also define an operator $\pi$ as $\pi\omega=\sum_p p\,\omega_p$.

We denote the Dirac operator both in the exterior bundle and Dirac bundle as ${D}$
\be\label{DO}
D = e^a\nabla_a\;,\qquad D=\gamma_* e^a\nabla_a=\gamma^a\nabla_a\,.
\ee
Here ${e^a\in\fiber{T}M\otimes\fiber{\Lambda} M}$ is a counterpart of ${\gamma_a}$ in the exterior algebra and $\nabla$ denotes the spinor covariant derivative. We also denote by ${X_a}$ the object dual to ${e^a}$ (see, e.g., Appendix of \cite{CarigliaEtal:2011} for details on the notation).

\subsection{First-order symmetry operators}
First-order operators commuting with the Dirac operator \eqref{DO} were recently studied in all dimensions
\cite{CarigliaEtal:2011}. Namely, we have the following result:
The most general first-order operator ${S}$ which commutes with the Dirac operator $D$, ${[D,S]=0}$, splits into the (Clifford) even and odd parts
\be\label{Scom}
  S=S_\even+S_\odd\,,
\ee
where
\ba
S_\even&=& K_{f_\odd} \equiv X^a\hook f_\odd\nabla_{\!a}
       + \frac{\pi-1}{2\pi}d f_\odd\,,\label{Kdef}   \\
S_\odd&=& M_{h_\even} \equiv e^a\wedge h_\even\nabla_{\!a}
       - \frac{\dm-\pi-1}{2(\dm-\pi)}\delta h_\even\;, \label{Mdef}
\ea
with ${f_\odd}$ being an inhomogeneous odd {\em Killing--Yano} form, and
${h_\even}$ being an inhomogeneous even {\em closed conformal Killing--Yano} form.

On the Dirac bundle these operators read (denoting by $K_{f_\odd}=\gamma_* K_{f_\odd}$ and $M_{h_\even}=\gamma_* M_{h_\even}$)
\begin{align}
&\begin{aligned}
  K_{f_\odd}&=\sum_{\text{${p}$ odd}}\frac1{(p{-}1)!}\Bigl[\gamma^{a_1\dots a_{p{-}1}}(f_p)^a_{\ a_1\dots a_{p{-}1}}\nabla_a\\
            &\qquad+\frac{1}{2(p{+}1)^2}\gamma^{a_1\dots a_{p{+}1}}(d f_p)_{a_1\dots a_{p{+}1}}\Bigr]\,.
\end{aligned}\\
&\begin{aligned}
  M_{h_\even}&=\sum_{\text{${p}$ even}}\frac1{p\,!}\,\Bigl[\gamma^{aa_1\dots a_{p}}(h_p)_{a_1\dots a_{p}}\nabla_a\\
             &\qquad-\frac{p(n{-}p)}{2(n{-}p{+}1)}\gamma^{a_1\dots a_{p{-}1}}(\delta h_p)_{a_1\dots a_{p{-}1}}\Bigr]\,,
\end{aligned}
\end{align}
where $p$-forms $f_p$ and $h_p$ (with ${f_\odd=\sum_{\text{${p}$ odd}}f_p}$ and ${h_\even=\sum_{\text{${p}$ even}}h_p}$) are odd Killing--Yano and even closed conformal Killing--Yano tensors, respectively. That is, they satisfy the following equations:
\ba
\nabla_a(f_p)_{a_1\dots a_p}&=&\frac{1}{p+1}(df_p)_{aa_1\dots a_p}\,,\label{KY}\\
\nabla_a(h_p)_{a_1\dots a_p}&=&-\frac{p}{n-p+1}g_{a[a_1}(\delta h_p)_{a_2\dots a_p]}\,.\label{CCKY_generic}\quad
\ea

In odd number of spacetime dimensions the Hodge duality of Killing--Yano tensors translates into the corresponding relation of symmetry operators $K$
and $M$. Namely, let $z$ be the Levi--Civita $\dm$--form satisfying
$z_{a_1\dots a_n} z^{a_1\dots a_n}= \dm !$.\footnote{%
Note that $\gamma_* (z)$ is the ordered product of all $\dm$ gamma matrices and in odd dimensions it is proportional to unit matrix. See also Sec.~\ref{ssc:csop}.
}
Then the Hodge dual of a $p$--form $\omega$ can be written as
\be   \label{eq:HodgeDual}
*\omega = (-1)^{(\dm-1)p + \left[\frac{p}{2}\right]} z \omega \, ,
\ee
and for the operators of type $K$ and $M$ it holds that
\be  \label{eq:M_to_K}
K_{z h}=(-1)^{\dm-1}zM_h\,,\quad
M_{z\!f} = (-1)^{\dm -1} z K_{\!f} \, ,
\ee
where $f$ is an odd KY form and $h$ an even CCKY form.

\section{Kerr-NUT-(A)dS spacetimes}\label{sec3}
We shall concentrate on the Dirac equation in general rotating Kerr-NUT-(A)dS spacetimes in all dimensions \cite{ChenEtal:2006cqg}. Slightly more generally we consider the most general {\em canonical metric} admitting a PCKY tensor \cite{HouriEtal:2007, KrtousEtal:2008} and intrinsically characterize separability of the massive Dirac equation in such a background.

The canonical metric is written as\footnote{We assume Euclidean signature of the metric. The physical signature could be obtained by a proper choice of signs of the metric function, a suitable Wick rotation of the coordinates and metric parameters, and a slight modification of various spinor-related conventions.}
\be  \label{metric}
{g}
  = \sum_{\mu=1}^N\biggl[ \frac{d x_{\mu}^{\;\,2}}{Q_\mu}
  +Q_\mu\Bigl(\,\sum_{j=0}^{N-1} \A{j}_{\mu}d\psi_j \Bigr)^{\!2}  \biggr]  + \eps S \Bigl(\,\sum_{j=0}^N \A{j}d\psi_j \Bigr)^{\!2}.
\ee
Here, coordinates $x_\mu\, (\mu=1,\dots,N)$ stand for the (Wick rotated) radial coordinate and longitudinal angles, and Killing
coordinates $\psi_k\; (k=0,\dots,N-1 +\eps)$ denote time and azimuthal angles associated with Killing vectors
${\KV{k}}$
\begin{equation}\label{KV}
\KV{k}=\cv{\psi_k}\;,\quad
\xi_{(k)}\equiv (\cv{\psi_k})^\flat\,.
\end{equation}
We have further defined\footnote{%
In what follows we assume no implicit summing over ${\mu,\nu,\dots}$ and ${j,k,l,m,\dots}$ indices. The explicit sums have, unless specifically indicated otherwise, ranges ${1,\dots,N}$ and ${0,\dots,N-1+\eps}$, respectively.
}
 the functions
(note that our sign convention for ${U_\mu}$ differs from the one in \cite{OotaYasui:2008})
\ba
Q_\mu&=&\frac{X_\mu}{U_\mu}\,,\quad U_{\mu}=\prod\limits_{\nu\ne\mu} (x_{\nu}^2-x_{\mu}^2)  \;,\quad S = \frac{-c}{\A{N}} \, ,\label{eq:UandS_def}\\
\A{k}_{\mu}&=&\hspace{-5mm}\!\!
    \sum\limits_{\substack{\nu_1,\dots,\nu_k\\\nu_1<\dots<\nu_k,\;\nu_i\ne\mu}}\!\!\!\!\!\!\!\!\!\!
    x^2_{\nu_1}\cdots\, x^2_{\nu_k}\;,\ \
\A{j} = \hspace{-5mm} \sum\limits_{\substack{\nu_1,\dots,\nu_k\\\nu_1<\dots<\nu_k}}\!\!\!\!\!\!
    x^2_{\nu_1}\cdots\, x^2_{\nu_k}\; .\label{eq:A_def}\quad
\ea
Functions $\A{k}_{\mu}$ and $\A{j}$ can be generated as follows:
\be
\begin{gathered}\label{Arel}
\prod_\nu \left(t- x_\nu^2 \right) = \sum_{j=0}^{N} (-1)^j \A{j} t^{N-j} \, , \\
\prod_{\substack{\nu\\\nu\neq \mu}} \left(t- x_\nu^2 \right) = \sum_{j=0}^{N} (-1)^j \A{j}_\mu t^{N-1-j} \, ,
\end{gathered}
\ee
and satisfy the important relations
\begin{equation}\label{AUrel}
\hspace*{-1mm}  \sum_{\mu}\! \frac{\A{i}_\mu}{U_\mu}{(-x_\mu^2)^{N{-}1{-}j}} = \delta^i_j\;,\quad
  \sum_{j}\!\frac{\A{j}_\mu}{U_\nu} {(-x_\nu^2)^{N{-}1{-}j}} = \delta^\nu_\mu\;.
\end{equation}

The quantities ${X_\mu}$ are functions of a single variable ${x_\mu}$, and $c$ is an arbitrary constant.
The vacuum (with a cosmological constant) black hole geometry is recovered by setting
\begin{equation}\label{BHXs}
  X_\mu = \sum_{k=\eps}^{N}\, c_{k}\, x_\mu^{2k} - 2 b_\mu\, x_\mu^{1-\eps} + \frac{\eps c}{x_\mu^2} \; .
\end{equation}
This choice of $X_\mu$ describes the most general known Kerr-NUT-(A)dS spacetimes in all dimensions \cite{ChenEtal:2006cqg}. The constant $c_N$ is proportional to the cosmological constant and the remaining constants are related to angular momenta,
mass and NUT parameters.

At points with ${x_\mu=x_\nu}$ with ${\mu\neq\nu}$ the coordinates are degenerate. We assume a domain where ${x_\mu\neq x_\nu}$ for ${\mu\neq\nu}$. In such a domain we can always order and rescale the coordinates in such a way that
\begin{equation}\label{coorord}
    x_\mu + x_\nu > 0 \quad\text{and} \quad x_\mu - x_\nu > 0 \quad \text{for}\;\mu<\nu\;.
\end{equation}
With this convention and assuming positive signature we have
\begin{equation}\label{eq:module}
    U_\mu = (-1)^{N{-}\mu} |U_\mu|\;,\quad X_\mu=(-1)^{N{-}\mu} |X_\mu|\;.
\end{equation}

We introduce the following orthonormal covector frame $\ef^a={\{\ef^\mu,\ef^{\hat\mu}, \ef^0\}}$,
\begin{equation}  \label{formframe}
\begin{gathered}
\ef^\mu = \frac{d x_{\mu}}{\sqrt{Q_\mu}}\;,\quad
  \ef^{\hat\mu} = \sqrt{Q_\mu} \sum_{j=0}^{N-1}\A{j}_{\mu}d\psi_j   \;, \\
E^0 = \sqrt{S} \sum_j \A{j}d\psi_j  \, ,
\end{gathered}
\end{equation}
and the dual vector frame $\ef_a={\{\ef_\mu,\ef_{\hat\mu}, \ef_0\}}$.
\begin{gather}
 \ef_\mu = \sqrt{Q_\mu}\cv{x_{\mu}}\;,\quad
 \ef_{\hat\mu} = \sqrt{Q_\mu}\sum_{j}\frac{(-x_\mu^2)^{N{-}1{-}j}}{X_\mu}\cv{\psi_j} \; , \nn\\
 E_0 = \frac{1}{\sqrt{S} \A{N}} \cv{\psi_N} \, , \label{vectfr}
\end{gather}
Note, that $E^0$ and $E_0$ are defined only in an odd dimension.
In this frame, the metric reads
\begin{equation}\label{diagmetric}
{g}= \sum_{\mu}\,
    \Bigl(\,\ef^\mu \otimes \ef^\mu
    + \ef^{\hat \mu} \otimes \ef^{\hat\mu}\,\Bigr) + \eps \ef^0 \otimes \ef^0
  \;,
\end{equation}
and the Ricci tensor is diagonal \cite{HamamotoEtal:2007},
\begin{equation}\label{Ric}
    \Ric = \sum_{\mu} r_\mu
    \Bigl(\,\ef^\mu \otimes \ef^\mu
    + \ef^{\hat \mu} \otimes \ef^{\hat\mu}\,\Bigr) + \eps r_0 \ef^0 \otimes \ef^0  \;,
\end{equation}
where
\ba \label{rmu}
    r_\mu &=& - \frac1{2x_\mu}\biggl[\sum_{\nu}
    \frac{x_\nu^2\bigl(x_\nu^{-1} \hat{X}_\nu\bigr){}_{,\nu}}{U_\nu} + \eps \sum_\nu \frac{\hat{X}_\nu}{U_\nu} \biggr]_{\!,\mu}\; , \nn \\
 r_0 &=& - \sum_\rho \frac{1}{x_\rho} \left( \sum_\sigma \frac{\hat{X}_\sigma}{U_\sigma} \right)_{\! , \rho} \, ,
\ea
and $\hat{X}_\mu = X_\mu - {\eps c}/{x_\mu^2} \, .$ For the Einstein space, polynomials \eqref{BHXs} lead to a constant value
${r_\mu}$.

The canonical metric \eqref{metric} possesses a hidden symmetry of the PCKY tensor \cite{KubiznakFrolov:2007}. In the basis \eqref{formframe} the PCKY 2-form reads
\begin{equation}\label{PCKY}
    h = \sum_{\mu=1}^N x^\mu\, \ef^\mu\wedge\ef^{\hat\mu}\;.
\end{equation}
This tensor generates the tower of closed conformal Killing--Yano $(2j)$-forms (note that this definition differs by the factorial from \cite{CarigliaEtal:2011}):
\begin{equation}\label{CCKY}
    \CCKY{j} = \frac1{j!}\,h^{\wedge j} = \frac1{j!}\, h\wedge\dots\wedge h\;,
\end{equation}
which in their turn give rise to Killing--Yano forms
\be\label{fj}
f^{(j)}=*h^{(j)}=(-1)^jz\CCKY{j}\,.
\ee
In the second equality we have used \eqref{eq:HodgeDual}.
Killing--Yano tensors \eqref{fj} `square to' 2nd-rank Killing tensors
\begin{equation}\label{KT}
\KT{j}=\sum_{\mu}\A{j}_\mu
\Bigl(\ef^\mu \otimes  \ef^\mu\! +\!
\ef^{\hat \mu}\otimes\ef^{\hat \mu}\Bigr) + \eps \A{j} \ef^0 \otimes  \ef^0   \;.
\end{equation}
Obviously, $\KT{0}$ coincides with the metric and hence it is a trivial Killing tensor which we include in our tower; so we take $j=0,\dots, N-1$.

The PCKY tensor $h$ generates also all the isometries \eqref{KV} of the spacetime. In particular, the {\em primary Killing vector} ${\PKV}=\KV{0}$ is
given by:
\be \label{PKV_indices}
  \PKV^a = \frac1{n-1}\nabla_{\!b}h^{ba} \,,
\ee
which can be written explicitly as
\be \label{PKV}
  \PKV = \sum_\mu \sqrt{Q_\mu} E_{\hat{\mu}} + \eps \sqrt{S} E_0 =\cv{\psi_0}\,.
\ee
It satisfies the important relation
\begin{equation}\label{nablaCCKY}
    -\frac1{n{-}2j{+}1}\,\delta\CCKY{j} = \PKV^\flat \wedge\CCKY{j{-}1}\;.
\end{equation}
In odd dimensions we also have
\be\label{cood}
\xi_{(N)}=(\partial_{\psi_N})^\flat=\sqrt{-c}*h^{(N)}=\sqrt{-c}f^{(N)}\,.
\ee

Let us finally mention that the explicit symmetries $\KV{k}$ and hidden symmetries $\KT{j}$ are responsible for complete integrability of the geodesic motion  as well as for separability of the Hamilton--Jacobi equation in spacetimes \eqref{metric}. Moreover, the corresponding operators
$\{(\KV{k})^a\nabla_a, \nabla_a (\KT{j})^{ab}\nabla_b\}$ form a complete set of commuting operators which intrinsically characterize separability of  the Klein--Gordon equation in these spacetimes \cite{SergyeyevKrtous:2008}.

\section{Complete set of Dirac symmetry operators}\label{sc:csop}

\subsection{Operators of the complete set}\label{ssc:csop}
The canonical spacetime \eqref{metric} admits a complete set of first order symmetry operators of the Dirac operator that are mutually commuting \cite{CarigliaEtal:2011}. These operators are determined by the tower of symmetries built from the PCKY tensor ${h}$. Namely, they are given by $({N}+\eps)$ KY 1-forms $\xi_{(k)}$, \eqref{KV}, and ${N}$ closed conformal Killing--Yano forms ${\CCKY{j}}$, \eqref{CCKY}. In the exterior algebra notation they read:
\be
    K_k \equiv K_{\xi_{(k)}} = X^a\hook \xi_{(k)}\nabla_{a} + \frac{1}{4}d\xi_{(k)} \, ,      \label{opsK}
\ee
for $k = 0, \dots, N-1+\eps$, and
\be
    M_j = M_{\CCKY{j}} \equiv e^a\wedge\CCKY{j}\nabla_{\!a} - \frac{n-2j}{2(n-2j+1)}\delta\CCKY{j}\, ,  \label{opsM}
\ee
for $j = 0, \dots, N-1$. Note that the operator $M_0$ corresponds to the Dirac operator, $M_0=D$.
It is the aim of this section to find an explicit representation of the action of these operators on the Dirac bundle.
As usual, we shall denote it by the same letter, i.e., we write ${K_j=\gamma_*K_j}$ and ${M_j=\gamma_*M_j}$.

Let us remark here that in odd dimensions, one has a `different choice' of operators commuting with the Dirac
operator---associated with (in this case odd) Killing--Yano tensors $f^{(j)}$, \eqref{fj}.
Using first relation \eqref{eq:M_to_K} one finds that
\be
K_{\!f^{(j)}}=(-1)^j z M_j\,.
\ee
Since in our representation (introduced below) we shall have $\gamma_*(z)=i^N$, i.e., a trivial matrix, we can without loss of generality
consider only operators $M_j$. (Operators $K_{\!f^{(j)}}=(-1)^j i^N M_j$ have the same eigenvectors.) In particular, due to
\eqref{cood}, we have the following identification:
\be\label{MNKN}
K_N=(-i)^N\!\sqrt{-c}\,M_N\,,
\ee
which shall be used in Sec.~\ref{sc:sep}.

\subsection{Representation of $\gamma$-matrices and spinors}\label{sc:DB}
In the canonical spacetimes \eqref{metric} the geometry determines a special frame $\ef^a$, \eqref{formframe}. This frame can be lifted to the frame ${\tht_E}$ in the Dirac bundle by demanding that the abstract gamma matrices ${\gamma^a}$ have constant components ${(\gamma^{a})^A{}_B}$, set to some special values.
It was a key observation of \cite{OotaYasui:2008} that these components can be chosen as a tensor product of ${N}$ 2-dimensional matrices. In other words, the special geometric structure of canonical spacetimes allows us to represent the fiber of the Dirac bundle as a tensor product of ${N}$ 2-dimensional spaces ${\fiber{S}}$, ${\fiber{D}M=\fiber{S}^N M}$ with gamma matrices adjusted to hidden symmetry. We use Greek letters ${\epsilon,\varsigma,\dots}$ for tensor indices in these 2-dimensional spaces and we use values ${\epsilon=\pm1}$ (or just ${\pm}$) to distinguish the components.

It means that we choose a frame ${\tht_E}$ in the Dirac bundle in a tensor product form:
\begin{equation}\label{Thetafr}
    \tht_E = \tht_{\epsilon_1\dots\epsilon_N}
      = \tht_{\epsilon_1}\otimes\dots\otimes\tht_{\epsilon_N}\;,
\end{equation}
where $\tht_+$ and $\tht_-$ form a frame in the 2-dimensional spinor space~${\fiber{S}}$. With such a choice we have a natural identification of Dirac indices ${E}$ with the multi-index ${\{\epsilon_1,\dots,\epsilon_N\}}$.

A generic 2-dimensional spinor can thus be written as $\chi = \chi^+ \tht_+ + \chi^- \tht_- = \chi^\epsilon \tht_\epsilon$, with components being two complex numbers ${\left( \begin{smallmatrix} \chi^+ \\ \chi^- \end{smallmatrix} \right)}$.
Similarly, the Dirac spinors ${\psi\in\fiber{D} M}$ can be written as $\psi = \psi^{\epsilon_1\dots\epsilon_N} \tht_{\epsilon_1\dots\epsilon_N}$ with ${2^N}$ components ${\psi^{\epsilon_1\dots\epsilon_N}}$.

Before we write down the gamma matrices in this frame, let us introduce some useful notation. Let ${\one}$, ${\iota}$, ${\sigma}$, and ${\hat\sigma}$ be the unit and respectively Pauli operators on ${\fiber{S}}$, i.e., their action is given in components by
\begin{equation}\label{sigmamatr}
\begin{gathered}
    (\one\,\chi)^\epsilon = \chi^\epsilon\;,\quad
    (\iota\,\chi)^\epsilon = \epsilon\,\chi^\epsilon\;,\\
    (\sigma\,\chi)^\epsilon = \chi^{-\epsilon}\;,\quad
    (\hat\sigma\,\chi)^\epsilon = -i\epsilon\,\chi^{-\epsilon}\;.
\end{gathered}
\end{equation}
In matrix form they are written as:
\begin{equation}\label{sigmamatrcomp}
\begin{gathered}
    \one^\epsilon{}_\varsigma \equiv
    \left(\begin{array}{cc}
        1 & 0 \\
        0 & 1 \\
      \end{array}\right)
    \;,\quad
    \iota^\epsilon{}_\varsigma \equiv
    \left(\begin{array}{cc}
        1 & 0 \\
        0 & -1 \\
      \end{array}\right)
    \;,\\
    \sigma^\epsilon{}_\varsigma \equiv
    \left(\begin{array}{cc}
        0 & 1 \\
        1 & 0 \\
      \end{array}\right)
    \;,\quad
    \hat\sigma{}^\epsilon{}_\varsigma \equiv
    \left(\begin{array}{cc}
        0 & -i \\
        i & 0 \\
      \end{array}\right)
    \;.
\end{gathered}
\end{equation}
These operators satisfy the standard relations
\begin{equation}\label{su2alg}
    \iota\sigma=-\sigma\iota=i\,\hat\sigma\;,\quad
    \sigma\hat\sigma=-\hat\sigma\sigma=i\,\iota\;,\quad
    \hat\sigma\iota=-\iota\hat\sigma=i\,\sigma\;.
\end{equation}
Next, for any linear operator ${\alpha\in\fiber{S}^1_1M}$ we denote by ${\alpha_{\lst{\mu}}\in\fiber{D}^1_1M}$ a linear operator on the Dirac bundle
\begin{equation}\label{alphaDB}
    \alpha_{\lst{\mu}} \equiv \one\otimes\dots\otimes\one\otimes\alpha\otimes\one\otimes\dots\otimes\one
\end{equation}
with ${\alpha}$ on the ${\mu}$-th place in the tensor product. Similarly, for mutually different indices ${\mu_1,\dots,\mu_j}$ we define
\begin{equation}\label{multialphaDB}
    \alpha_{\lst{\mu_1\dots\mu_j}} \equiv \alpha_{\lst{\mu_1}}\otimes\dots\otimes\alpha_{\lst{\mu_j}}\;.
\end{equation}

Now we can finally write down the abstract gamma matrices with respect to the frame $\ef_a=\{{\ef_\mu,\,\ef_{\hat\mu}}, \ef_0\}$ chosen in tangent space,
\begin{equation}\label{gammamatrmuhatmu}
\begin{gathered}
    \gamma^\mu = \iota_{\lst{1\dots\mu{-}1}}\sigma_{\lst{\mu}}\;,\quad
    \gamma^{\hat\mu} = \iota_{\lst{1\dots\mu{-}1}}\hat\sigma_{\lst{\mu}}\, , \\
    \gamma^{0} = \iota_{\lst{1 \dots N}} \, ,
\end{gathered}
\end{equation}
where $\gamma^{0}$ is defined only in odd dimension. This definition essentially fixes the relation of the spinor frame ${\tht_E}$ to the frame in the tangent space. It is straightforward to check that the matrices \eqref{gammamatrmuhatmu} satisfy the property \eqref{gamgamg}.

In components, the action of these matrices on a spinor $\psi = \psi^{\epsilon_1\dots\epsilon_N} \tht_{\epsilon_1\dots\epsilon_N}$ is given as
\begin{equation}\label{gammaaction}
\begin{aligned}
    (\gamma^\mu\psi)^{\epsilon_1\dots\epsilon_N} &=
     \Bigl(\prod_{\nu =1}^{\mu-1} \epsilon_\nu\Bigr)\,\psi^{\epsilon_1\dots(-\epsilon_\mu)\dots\epsilon_N}\;,\\
    (\gamma^{\hat\mu}\psi)^{\epsilon_1\dots\epsilon_N} &=
     -i\epsilon_\mu\,\Bigl(\prod_{\nu=1}^{\mu-1} \epsilon_\nu\Bigr)\psi^{\epsilon_1\dots(-\epsilon_\mu)\dots\epsilon_N}\;, \\
    (\gamma^{0}\psi)^{\epsilon_1\dots\epsilon_N} &=
     \Bigl(\prod_{\nu =1}^{N} \epsilon_\nu\Bigr)\,\psi^{\epsilon_1\dots \epsilon_N}\;.
\end{aligned}
\end{equation}
We shall also use the relations
\be\label{relG}
\gamma^{\hat \mu}=-i\iota_{\lst{\mu}}\gamma^\mu\,,\quad  \gamma^{\mu \hat \mu}=i \iota_{\lst{\mu}}\,
\ee
and the fact that $\gamma_*(z)=\gamma^{\mu_1 \dots \mu_N\hat \mu_1\dots \hat \mu_N\,0}=i^N\one$\,.

\subsection{Explicit form of the operators}

Symmetry operators determined by Killing vectors are, in general, equivalent to the Lie derivative lifted from the tangent bundle to the Clifford or Dirac bundles. Thanks to \eqref{KV} we can thus write
\begin{equation}\label{opKexpl}
    K_{k} = \lied_{\KV{k}} = \cd{\psi_k}\;,
\end{equation}
where ${\cd{\psi_k}}$ is a partial derivative along ${\psi_k}$ which acts only on the components of the spinor in the frame ${\tht_E}$ described above. That is, let $\chi=\chi^E \tht_E$ be a spinor, then $K_{\!k}\chi={\cd{\psi_k}{\chi}=\frac{\partial\chi^E}{\partial \psi_k}\tht_E}$ (cf.\ Note~\ref{nt:thtder}).

The operators ${M_j}$, \eqref{opsM}, must be lifted to the Dirac bundle by using \eqref{cleaiso}. Let's start with expressing the action of a form given by ${\alpha\wedge\CCKY{j}}$ with ${\alpha}$ being a 1-form:
\begin{equation}\label{ClfaCCKY1}
\begin{split}
  &\gamma_*\bigl(\alpha\wedge\CCKY{j}\bigr)\\
    &\;= \frac1{j!(2j+1)!}\,
       \left( \alpha \wedge h \wedge \dots \wedge h \right)_{a_0 a_1 \dots a_2j} \,\gamma^{a_0a_1\dots a_{2j}}\\
    &\;= \frac1{j!\, 2^j}\, \alpha_{a_0}h_{a_1a_2}\dots h_{a_{2j{-}1}a_{2j}} \,\gamma^{a_0a_1\dots a_{2j}}\\
    &\;= \frac1{j!\, 2^j} \!\!\!\!\!\sum_{\substack{a_0,a_1,\dots,a_{2j} \\a_i\text{ all different}}}\!\!\!\!\!\!\!
       \alpha_{a_0}h_{a_1a_2}\dots h_{a_{2j{-}1}a_{2j}} \,\gamma^{a_0}\gamma^{a_1}\!\dots \gamma^{a_{2j}}.
\end{split}\raisetag{13ex}
\end{equation}
In the last equality we assumed that indices ${a_i}$ correspond to the orthonormal frame ${E_\mu,\,E_{\hat\mu}}, \, E_0$ which imply that gamma matrices ${\gamma^a}$ with different indices anticommute. In such frame, however, the PCKY tensor ${h}$ has only nonzero components ${h_{\mu\hat\mu}=-h_{\hat\mu\mu}=x_\mu}$. We can thus write
\begin{equation}\label{ClfaCCKY2}
\begin{split}
  &\gamma_*\bigl(\alpha\wedge\CCKY{j}\bigr)\\
    &= \frac1{j!} \sum_{\mu}\!\!\!
       \sum_{\substack{\mu_1,\dots,\mu_j\\\mu_i\text{ different}\\\mu_i\neq\mu}}\!\!\!\!\!\!\!
       \bigl(\alpha_\mu \gamma^\mu +\alpha_{\hat\mu}\gamma^{\hat\mu}\bigr) x_{\mu_1}\dots x_{\mu_j} \,
       \gamma^{\mu_1\hat \mu_1\dots \mu_j\hat\mu_j}\\
   &\ \ \  + \eps\frac1{j!}\alpha_0\gamma^0 \!\!\!\!\! \sum_{\substack{\mu_1,\dots,\mu_j\\\mu_i\text{ different}}}\!\!\!\!
       x_{\mu_1}\dots x_{\mu_j} \, \gamma^{\mu_1\hat \mu_1\dots \mu_j\hat\mu_j}\\
    &= i^j \sum_\mu \Bigl(\sum_{\substack{\mu_1,\dots,\mu_j\\\mu_1<\dots<\mu_j\\\mu_i\neq\mu}}
       \!\!\!\!x_{\mu_1}\dots x_{\mu_j} \, \iota_{\lst{\mu_1\dots\mu_j}}\Bigr)
       \bigl(\alpha_\mu \gamma^\mu +\alpha_{\hat\mu}\gamma^{\hat\mu}\bigr) \\
     &\ \ \ +\eps i^j\Bigl(\sum_{\substack{\mu_1,\dots,\mu_j\\\mu_1<\dots<\mu_j}}
       x_{\mu_1}\dots x_{\mu_j} \, \iota_{\lst{\mu_1\dots\mu_j}}\Bigr)\alpha_0\gamma^0
        \;,
\end{split}\raisetag{8ex}
\end{equation}
In the last equality we have used \eqref{relG} and the symmetry of the summands with respect to permutation of the ${\mu_i}$ indices.
The result can be rewritten as
\begin{equation}\label{ClfaCCKY}
  \gamma_*\bigl(\alpha\!\wedge\!\CCKY{j}\bigr) = i^j \!\sum_\mu \Bo{j}_\mu
       \bigl(\alpha_\mu \gamma^\mu\! +\!\alpha_{\hat\mu}\gamma^{\hat\mu} \bigr)\!+\!\eps i^j \Bo{j} \alpha_{0} \gamma^{0} \;,
\end{equation}
where we introduced a spinorial analogue of functions ${\A{j}_\mu}$ and ${\A{j}}$,  \eqref{eq:A_def}--\eqref{Arel}, given by
\ba
\Bo{k}_{\mu}&=&\hspace{-3mm}
    \sum\limits_{\substack{\nu_1,\dots,\nu_k\\\nu_1<\dots<\nu_k,\;\nu_i\ne\mu}}\!\!\!\!\!
    \iota_{\lst{\nu_1}}x_{\nu_1}\cdots\ \iota_{\lst{\nu_k}}x_{\nu_k} \;,\nonumber\label{B_mu_def}\\
  \Bo{k}&=& \hspace{-1mm}\sum\limits_{\substack{\nu_1,\dots,\nu_k\\\nu_1<\dots<\nu_k}}\!\!\!\!\!
    \iota_{\lst{\nu_1}}x_{\nu_1}\cdots\ \iota_{\lst{\nu_k}}x_{\nu_k} \;.  \label{Bdef}
\ea
These functions are elementary symmetric functions of $\left\{ \iota_{\lst{\nu}}x_\nu \right\}$ and $\left\{ \iota_{\lst{\nu}}x_\nu \right\}_{\nu\neq \mu}$ respectively:
\be
\begin{gathered}\label{Brel}
\prod_\nu \left(t- \iota_{\lst{\nu}}x_\nu \right) = \sum_{j=0}^{N} (-1)^j \Bo{j} t^{N-j} \, , \\
\prod_{\substack{\nu\\\nu\ne\mu}} \left(t- \iota_{\lst{\nu}}x_\nu \right) = \sum_{j=0}^{N} (-1)^j \Bo{j}_\mu t^{N-1-j} \, . \\
\end{gathered}
\ee
Relations including $\Bo{k}_{\mu}$ and $\Bo{k}$ can be formally obtained from the corresponding expressions valid for $\A{k}_{\mu}$ and $\A{k}$ by using a simple
rule
\be\label{subs}
A\leftrightarrow B \quad \Leftrightarrow \quad x_\mu^2\leftrightarrow \iota_{\lst{\mu}}x_{\mu}\,.
\ee
In particular, we can introduce
the analogue of functions $U_\mu$, \eqref{eq:UandS_def}, by
\be
 \Vo_{\mu}=\prod\limits_{\substack{\nu\\\nu\ne\mu}} (\iota_{\lst{\nu}} x_{\nu}-\iota_{\lst{\mu}}x_{\mu})\;.\label{Vdef}
 \ee
 and derive the relations analogous to Eq.~\eqref{AUrel}
 \begin{equation}\label{BVrel}
\begin{gathered}
\sum_{\mu} \frac{\Bo{i}_\mu}{\Vo_\mu}\,
{\bigl(-\iota_{\lst{\mu}}x_\mu\bigr)^{N{-}1{-}j}} =
  \delta^i_j\;,\\
\sum_{j}\frac{\Bo{j}_\mu}{\Vo_\nu}\, {\bigl(-\iota_{\lst{\nu}}x_\nu\bigr)^{N{-}1{-}j}} = \delta^\nu_\mu\;.
\end{gathered}
\end{equation}
Additional important relations regarding quantities $\Bo{k}_{\mu}$ and $\Bo{k}$ are gathered in App.~\ref{apx:ident}.

After this preliminary work we are ready to find the action of the operators $M_j$ \eqref{opsM}.
Let us start with the second term in \eqref{opsM}.
Using \eqref{nablaCCKY}, formula \eqref{ClfaCCKY}, the explicit expression for the primary Killing vector \eqref{PKV}, and the first relation \eqref{relG},
we find
\begin{equation}\label{0ordcoef}
\begin{split}
&\gamma_*\Bigl(-\frac{n{-}2j}{2(n{-}2j{+}1)}\delta\CCKY{j}\Bigr)
    =\frac{1}{2}(n{-}2j)\gamma_*\Bigl(\PKV^\flat\wedge\CCKY{j{-}1}\Bigr)\\
&\quad=-i^{j} (N\!-\!j\!+\!\varepsilon/2) \Bigl(\,
    \sum_\mu\!\sqrt{Q_\mu}\,\Bo{j{-}1}_\mu\iota_{\lst{\mu}}\gamma^{\mu} \\
    &\mspace{240mu}+  i \eps \sqrt{S} \Bo{j-1} \gamma^0 \Bigr)\, .
\end{split}\raisetag{10ex}
\end{equation}
Next, we want to find the expression for
$\gamma_*(e^a\wedge \CCKY{j}\nabla_a)$, where the spin derivative
with respect to the chosen frame $\ef^a$ is
\begin{equation}\label{covdfr}
    \nabla_a = \eth_a + \frac14\omega_{abc}\gamma^b\gamma^c\;.
\end{equation}
Here  ${\eth_a}$ is the derivative acting only on components of the spinor\footnote{\label{nt:thtder}
The derivative ${\eth_a}$ annihilates the frames ${\ef^a\equiv\{\ef^\mu,\ef^{\hat\mu}, \ef^{0}\}}$ and ${\tht_E}$, ${\eth_a\ef^\mu=\eth_a\ef^{\hat\mu}=\eth_a\ef^0=0}$, ${\eth_a{\tht_E}=0}$. It thus acts just on the components, ${\eth_a{\alpha}=\bigl(\partial_a\alpha_{b}\bigr)\ef^b}$ and ${\eth_a{\chi}=\bigl(\partial_a\chi^E\bigr)\tht_E}$. The connection coefficients are defined as ${\nabla_{\!a}\ef^b=-\omega{}_a{}^b{}_c \ef^c }$.}
and the connection coefficients ${\omega_{abc}}$ are listed in App. \ref{apx:spincon}.
Using \eqref{ClfaCCKY} we have
\begin{gather}\label{1ordcoef}
    \gamma_*(e^\mu\wedge \CCKY{j}) = i^j\Bo{j}_\mu\gamma^\mu\;,\quad
    \gamma_*(e^{\hat\mu}\wedge \CCKY{j}) = i^j\Bo{j}_\mu\gamma^{\hat\mu} \;, \nn \\
    \gamma_*(e^0\wedge \CCKY{j}) = i^j \Bo{j}\gamma^{0} \;.
\end{gather}
Hence, the derivative term can be expressed, with help of \eqref{vectfr}, in terms of partial derivatives
\begin{equation}\label{ethder}
\begin{split}
& \gamma_*(e^a\wedge\CCKY{j})\,\eth_a \\
& = i^j \sum_\mu \!\sqrt{Q_\mu} \Bo{j}_\mu
     \Biggr[ \cd{x_\mu} - \frac{i\iota_{\lst{\mu}}}{X_\mu}\sum_k (-x_\mu^2)^{N{-}1{-}k}\cd{\psi_k}\Biggr]
    \gamma^\mu \\
& \mspace{185mu} + \eps\, i^j \frac{\Bo{j}}{\sqrt{S}\A{N}} \cd{\psi_N} \gamma^0  \; .
\end{split}\raisetag{4.5ex}
\end{equation}
Moreover, using the explicit form of the connection coefficients, we find
\begin{equation}\label{conctterm}
\begin{split}
& \frac14\gamma_*(e^a\wedge\CCKY{j})\,\omega_{abc}\gamma^b\gamma^c  =\\
&i^j\sum_\mu\!\!\sqrt{Q_\mu}
       \Biggl(\!\frac{X_\mu'}{4X_\mu}\Bo{j}_\mu \!+\!\!\!
\sum_{\substack{\nu\\\nu\neq\mu}} \frac{x_\mu\!+\!\iota_{\lst{\mu\nu}}x_\nu}{x_\mu^2\!-\!x_\nu^2}
         \bigl(\Bo{j}_\nu\!-\!\frac12\Bo{j}_\mu\bigl)\!  \Biggr)\gamma^\mu \\
&\quad + \eps\, i^j\sum_\mu \Biggl(\frac{\sqrt{Q_\mu}}{2x_\mu}\,\Bo{j}\gamma^\mu\!+\!
\frac{i\sqrt{S}}{2x_\mu}\bigl(\Bo{j}\!-\!2\Bo{j}_\mu\bigr)\iota_{\lst{\mu}}\gamma^0 \Biggr)\,.
\end{split}\raisetag{21ex}
\end{equation}
Putting all three terms \eqref{0ordcoef}, \eqref{ethder} and \eqref{conctterm} together and using the identities \eqref{dveid} and \eqref{eq:Bid} we derive our final form for the operators $M_j$
\begin{equation}\label{Mopexpl}
\begin{split}
M_j
&= i^j \!\sum_\mu\! \sqrt{Q_\mu} \Bo{j}_\mu\Biggl(
    \cd{x_\mu}+\frac{X_\mu'}{4X_\mu} +  \frac12 \sum_{\substack{\nu\\\nu\neq\mu}} \frac1{x_\mu\!-\!\iota_{\lst{\mu\nu}}x_\nu}  \\
&- \frac{i\iota_{\lst{\mu}}}{X_\mu}\!\sum_k (-x_\mu^2)^{N{-}1{-}k}\cd{\psi_k}
    +\frac{\eps}{2x_\mu}\Biggr)\gamma^\mu   \\
&+ \eps\, i^{j+1} \frac{\sqrt{S}}{2} \Biggl[
   \Bo{j-1}\!-\!\Bo{j}\Bigl(\frac{2}{ic}\cd{\psi_N}\!+\!\sum_\mu\! \frac{1}{\iota_{\lst{\mu}} x_\mu}\Bigr)  \Biggr] \gamma^0 \, .
\end{split}\raisetag{13.5ex}
\end{equation}

\section{R-separability of the Dirac equation}\label{sc:sep}
Now we can formulate the main result: the commuting symmetry operators ${K_k}$ and ${M_j}$ have common spinorial eigenfunctions ${\psi}$
\begin{align}
    K_k \psi &= i\,\psc{k}\psi\;,\label{eigenfcK}\\
    M_j \psi &= \chc{j}\psi\;,\label{eigenfcM}
\end{align}
which can be found in the tensorial R-separated form
\begin{equation}\label{tensRsep}
    \psi = R\,\exp\bigl({\textstyle i\sum_k\psc{k}\psi_{k}}\bigr)\,
           \bigotimes_\nu \chi_\nu\;,
\end{equation}
where $\left\{\chi_\nu \right\}$ is an $N$-tuple of 2-dimensional spinors and ${R}$ is the (Clifford bundle)-valued prefactor\footnote{Note that thanks to our convention \eqref{coorord}, the operators under square root in ${R}$ are positively definite.}
\begin{equation}\label{Phidef}
  R = \prod_{\substack{\kappa,\lambda\\\kappa<\lambda}}
    \Bigl(x_\kappa+\iota_{\lst{\kappa\lambda}}x_\lambda\Bigr)^{-\frac12}\;.
\end{equation}
As part of the separation ansatz we ask that $\chi_\nu$ depends only on the variable ${x_\nu}$, $\chi_\nu=\chi_\nu(x_\nu)$. (Hence we have ${\cd{\psi_k}\chi_\nu=0}$ and ${\cd{x_\mu}\chi_\nu=0}$ for ${\nu\neq\mu}$.)

In this section, we are going to show that Eqs. \eqref{eigenfcK} and \eqref{eigenfcM} are satisfied if and only if the spinors $\chi_\nu$ satisfy the ordinary differential equations \eqref{chieq} below. These equations are equivalent to the conditions found in \cite{OotaYasui:2008}.

Let us first note that rewriting the tensorial R-separability ansatz \eqref{tensRsep} in terms of components, we recover the separated solution of the massive Dirac equation (equivalent to \eqref{eigenfcM} with ${j=0}$)  which was found in \cite{OotaYasui:2008}:
\begin{equation}\label{compsep}
    \psi^{\epsilon_1\dots\epsilon_N} =
      \phi_{\epsilon_1\dots\epsilon_N}
      \exp\bigl({\textstyle i\sum_k\psc{k}\psi_{k}}\bigr)
      \prod_\nu  \chi_\nu^{\epsilon_\nu} \;.
\end{equation}
Here, ${\phi_{\epsilon_1\dots\epsilon_N}}$ is a diagonal element of the prefactor ${R}$,
\begin{equation}\label{phicompdef}
    \phi_{\epsilon_1\dots\epsilon_N} =
    \prod_{\substack{\kappa,\lambda\\\kappa<\lambda}}
    \Bigl(x_\kappa+\epsilon_\kappa\epsilon_\lambda x_\lambda\Bigr)^{-\frac12}\;.
\end{equation}

To derive the announced results we shall work directly with the tensorial multiplicative ansatz \eqref{tensRsep} and only in the end we shall make contact with the
work of \cite{OotaYasui:2008} by finding an equation for the components of each ${\chi_\nu}$.

The spinor ${\psi}$ given by \eqref{tensRsep} satisfies \eqref{eigenfcK}. To show \eqref{eigenfcM}, we need to calculate $M_j\psi$, with $M_j$ given by \eqref{Mopexpl}.
We have
\begin{equation}\label{Mjpsi1}
\begin{split}
&M_j \psi=i^j \exp\bigl({\textstyle i\sum_k\psc{k}\psi_{k}}\bigr)
  \Biggl[\!\sum_\mu\! \sqrt{Q_\mu} \Bo{j}_\mu\times  \\
& \times\Biggl(
    \cd{x_\mu}{+}\frac{X_\mu'}{4X_\mu}
    {+}\frac12 \sum_{\substack{\nu\\\nu\neq\mu}} \frac1{x_\mu{-}\iota_{\lst{\mu\nu}}x_\nu}
    {+}\frac{\tilde \Psi_\mu}{X_\mu}\iota_{\lst{\mu}}
    {+}\frac{\eps}{2x_\mu}\Biggr)\gamma^\mu  \\
&+ \eps\, \frac{i\sqrt{S}}{2}
   \Biggl(\! \Bo{j\!-\!1}\!\!-\!\Bo{j}\Bigl(\frac{2 \tilde \Psi_N}{c}
   \!+\!\sum_\mu\! \frac{\iota_{\lst{\mu}}}{x_\mu}\Bigr)  \Biggr) \gamma^0 \!\Biggr]\,R\bigotimes_\nu  \chi_\nu\;,\\
\end{split}\raisetag{22.7ex}
\end{equation}
where we have performed the derivative with respect to angles ${\psi_k}$ and introduced the functions of one variable ${\psf{\mu}} $ given by
\begin{equation}\label{psfdef}
    \psf{\mu} = \sum_k \psc{k} (-x_\mu^2)^{N{-}1{-}k}\;.
\end{equation}

Let us concentrate now on the derivatives of the prefactor ${R}$. Using Eq. \eqref{derPhi}
and relation \eqref{eq:gamma_mu_phi} we can bring the operator $R$ to the front to get
\begin{equation}\label{Mjpsi2}
\begin{split}
&M_j\psi =i^j \exp\bigl({\textstyle i\sum_k\!\psc{k}\psi_{k}}\bigr)\,R\\
&\times\!\Biggl[\sum_\mu \!\frac{\sqrt{|X_\mu|}}{\Vo_{\mu}}
  \bigl(-\! \iota_{\lst{\mu}} \bigr)^{\!N\!-\!\mu}\Bo{j}_\mu\\
&\qquad\quad\quad\times\Bigl( \cd{x_\mu}+\frac{X_\mu'}{4X_\mu}
  +\frac{\tilde \Psi_\mu}{X_\mu}\iota_{\lst{\mu}}
  +\frac{\eps}{2x_\mu}\Bigr)\,\sigma_{\lst{\mu}} \\
&\quad+\! \eps\, \frac{i\sqrt{S}}{2} \Bigl(\! \Bo{j\!-\!1}
  {-}\Bo{j}\Bigl(\frac{2 \tilde \Psi_N}{c}{+}\sum_\mu\! \frac{\iota_{\lst{\mu}}}{x_\mu}\Bigr)
  \Bigr)\, \gamma^0 \Biggr]
  \bigotimes_\nu \chi_\nu\,.
\end{split}\raisetag{16ex}
\end{equation}
This expression is to be compared with
\be\label{msi3}
\chc{j}\psi= \chc{j}\exp\bigl({\textstyle i\sum_k\psc{k}\psi_{k}}\bigr)R
           \bigotimes_\nu \chi_\nu \,.
\ee

To simplify the following expressions we
introduce the functions $\chf{\nu}$ of a single variable ${x_\nu}$:
\be  \label{eq:chf_mu_definition}
\chf{\nu} =  \sum_{j} (-i)^j \chc{j} \left( -\iota_{\lst{\nu}} x_\nu \right)^{N-1-j} \, .
\ee
In odd dimensions the constant $\chc{N}$, defined by $M_N\psi=\chc{N}\psi$, is not independent.
In fact, using Eq. \eqref{MNKN} and \eqref{eigenfcK}, \eqref{eigenfcM}, we have
\be\label{chiN}
\chc{N}=\frac{i^{N+1}}{\sqrt{-c}}\Psi_N \, .
\ee

We are now ready to derive the differential equations for $\chi_\nu$ so that \eqref{eigenfcM} are satisfied.
We can cancel the common $\exp\bigl({\textstyle i\sum_k\psc{k}\psi_{k}}\bigr) R$ prefactor in \eqref{Mjpsi2} and
\eqref{msi3}  (in the coordinate domain we are using the operator $R$ is never zero on any spinor), multiply both equations by $(-i)^j \left( -\iota_{\lst{\nu}} x_\nu \right)^{N-1-j}$ and sum over $j$ to obtain
\begin{equation}\label{predpos}
\begin{split}
&\chf{\nu} \bigotimes_\kappa  \chi_\kappa=\\
&=\Biggl[\!\sqrt{|X_\nu|}\bigl(-\! \iota_{\lst{\nu}} \bigr)^{\!N\!-\!\nu}
  \Bigl(\cd{x_\nu}{+}\frac{X_\nu'}{4X_\nu}
  {+}\frac{\tilde \Psi_\nu}{X_\nu}\iota_{\lst{\nu}}
  {+}\frac{\eps}{2x_\nu}\Bigr)\sigma_{\lst{\nu}}\\
&\mspace{100mu}- \eps\, \frac{i\sqrt{S}}{2x_\nu^2} \Bo{N}\gamma^0 \Biggr]\,
  \bigotimes_\kappa  \chi_\kappa\,\,,
\end{split}\raisetag{5ex}
\end{equation}
where we have used the latter equation \eqref{BVrel} and identities
\eqref{eq:sum_Boj} and \eqref{eq:sum_Boj-1}. Using further the formula
\be\label{Sid}
 \gamma^0 \sqrt{S} = \frac{\sqrt{-c}}{\Bo{N}} \, ,
\ee
we can rewrite Eq. \eqref{predpos} as
\begin{equation}\label{pos}
\begin{split}
&\Biggl[\!\sqrt{|X_\nu|}\bigl(-\! \iota_{\lst{\nu}} \bigr)^{\!N\!-\!\nu}
\Bigl(
    \cd{x_\nu}{+}\frac{X_\nu'}{4X_\nu}
    {+}\frac{\tilde \Psi_\nu}{X_\nu}\iota_{\lst{\nu}}
    {+}\frac{\eps}{2x_\nu}\Bigr)\sigma_{\lst{\nu}} \\
&\mspace{100mu}- \eps\, \frac{i\sqrt{-c}}{2x_\nu^2} -\chf{\nu}\Biggr]\,\bigotimes_\kappa \chi_\kappa=0\,.
\end{split}\raisetag{5ex}
\end{equation}
We finally note that the operators act only on the $\chi_\nu$ spinor, leaving invariant all the other spinors in the tensor product. So we are left with the following ordinary differential equation for each spinor $\chi_\nu$:
\begin{equation}\label{chieq}
\begin{split}
&\Biggl[\Bigl(
    \frac{d}{dx_\nu}+\frac{X_\nu'}{4X_\nu}
    +\frac{\tilde \Psi_\nu}{X_\nu}\iota_{\lst{\nu}}
    +\frac{\eps}{2x_\nu}\Bigr)\,\sigma_{\lst{\nu}} \\
&\mspace{70mu}- \,\frac{\bigl(- \iota_{\lst{\nu}} \bigr)^{\!N\!-\!\nu}}{\sqrt{|X_\nu|}}
   \Bigl(\eps \frac{i\sqrt{-c}}{2x_\nu^2} +\chf{\nu}\Bigr)\Biggr]\,\chi_\nu=0\,.
\end{split}
\end{equation}

To make contact with the formalism of \cite{OotaYasui:2008} we redefine ${\chi_\nu}$ in an odd dimension by a suitable rescaling,
\begin{equation}\label{rescchi}
  \tilde\chi_\nu = (x_\nu)^{\frac\eps2}\chi_\nu\;.
\end{equation}
Taking the ${\varsigma}$-component of the spinorial equation \eqref{chieq} we then get
\begin{equation}\label{chieq_proj}
\begin{split}
&\Bigl(
    \frac{d}{d x_\nu}+\frac{X_\nu'}{4X_\nu}
    -\varsigma\frac{\tilde \Psi_\nu}{X_\nu}\Bigr)\,\tilde\chi_\nu^{-\varsigma} \\
&\mspace{100mu} - \frac{\bigl(-\varsigma\bigr)^{\!N\!-\!\nu}}{\sqrt{|X_\nu|}}
   \Bigl(\eps \frac{i\sqrt{-c}}{2x_\nu^2} +\chf{\nu}\Bigr)\,\tilde\chi_\nu^{\varsigma}=0\,.
\end{split}
\end{equation}
For each $\nu$, these are two coupled ordinary differential equations for components ${\tilde\chi_\nu^+}$ and ${\tilde\chi_\nu^-}$, which can be easily decoupled by substituting one into another.

It can be checked that these are equivalent to the differential equations given in \cite{OotaYasui:2008} with a proper identification of coefficients.
In particular for the eigenvalue of the Dirac equation the identification is $q_{N-1} = \chc{0}$, as expected.

\section{Standard separability}\label{sc:sep2}
In this section we shall comment on how to achieve the standard tensorial separability, without the prefactor $R$.
For this purpose it is first instructive to prove directly the commutativity of operators $M_j$. This will give us a hint how to `upgrade' our representation
to achieve the standard tensorial separability.

\subsection{Direct proof of commutativity}
Let us start from the expression for $M_j$ \eqref{Mopexpl}  and apply the identity  \eqref{Sid} and \eqref{id15} , to obtain
\begin{equation}\label{Mopexpl2}
\begin{split}
M_j
&= i^j \!\sum_\mu\! \sqrt{Q_\mu} \Bo{j}_\mu\Biggl(
    \cd{x_\mu}+\frac{X_\mu'}{4X_\mu} +  \frac12 \sum_{\substack{\nu\\\nu\neq\mu}} \frac1{x_\mu\!-\!\iota_{\lst{\mu\nu}}x_\nu}  \\
&- \frac{i\iota_{\lst{\mu}}}{X_\mu}\!\sum_k (-x_\mu^2)^{N{-}1{-}k}\cd{\psi_k}
    +\frac{\eps}{2x_\mu}\Biggr)\gamma^\mu   \\
&- \frac{1}{2}\eps\, i^{j+1} \sqrt{-c} \sum_\mu\frac{\Bo{j}_\mu}{\Vo_\mu}\left(\frac{1}{x_\mu^2}
+\frac{2}{ic}\frac{1}{\iota_{\lst{\mu}} x_\mu}\cd{\psi_N}\right) \, .
\end{split}\raisetag{13.5ex}
\end{equation}
In order to prove commutativity of these operators we introduce new `auxiliary' operators
\be\label{RMR}
\tilde M_j\equiv R^{-1}M_jR\,,
\ee
with $R$ given by \eqref{Phidef}.
Then, obviously, if
\be\label{com}
[\tilde M_j, \tilde M_k]=R^{-1}[M_j,M_k]R=0\,,
\ee
the same is true for operators without tilde.

We calculate
\begin{equation}\label{Mopexpl3}
\begin{split}
&M_jR
= i^j \!\sum_\mu\! \sqrt{Q_\mu} \Bo{j}_\mu R R^{-1}\gamma^\mu R\\
&\;\;\times \Biggl(
    \cd{x_\mu}+\frac{X_\mu'}{4X_\mu}+\frac{\eps}{2x_\mu}
    + \frac{i\iota_{\lst{\mu}}}{X_\mu}\!\sum_k (-x_\mu^2)^{N{-}1{-}k}\cd{\psi_k}
    \Biggr)   \\
&\;\;- \frac{1}{2}\eps\, i^{j+1} \sqrt{-c} \sum_\mu\frac{\Bo{j}_\mu}{\Vo_\mu}\left(\frac{1}{x_\mu^2}
+\frac{2}{ic}\frac{1}{\iota_{\lst{\mu}} x_\mu}\cd{\psi_N}\right)R \, .
\end{split}\raisetag{21.5ex}
\end{equation}
where  we have used \eqref{derPhi}.
The first $R$ on the right hand side can be brought to the front whereas for the product $R^{-1}\gamma^\mu R$ we use \eqref{eq:gamma_mu_phi}.
So we get
\be \label{tildeMj}
\tilde M_j =i^j\sum_\mu \frac{\Bo{j}_\mu}{ \Vo_{\mu}}\tilde M_\mu \,,
\ee
where the operators
\ba
\tilde M_\mu&=&\sqrt{|X_\mu|}\Biggl(
    \cd{x_\mu}+\frac14\frac{X_\mu'}{X_\mu} +  \frac{\eps}{2x_\mu}  \nn\\
      &&  - \frac{i\iota_{\lst{\mu}}}{X_\mu}\!\sum_k (-x_\mu^2)^{N{-}1{-}k}\cd{\psi_k} \Biggr)
\left( - \iota_{\lst{\mu}}\right)^{N-\mu} \sigma_{\lst{\mu}}\,,\nn\\
&&- \frac{i}{2}\eps\,\sqrt{-c}\left(\frac{1}{x_\mu^2}
+\frac{2}{ic}\frac{1}{\iota_{\lst{\mu}} x_\mu}\cd{\psi_N}\right) \,
\ea
act only on spinor $\chi_\mu$ and hence $[\tilde M_\mu,\tilde M_\nu]=0$.
Using \eqref{BVrel} we can invert the relation \eqref{tildeMj},
\be\label{tildeMmu}
\tilde M_\mu=\sum_{j=0}^{N-1}(-i)^j(-\iota_{\lst{\mu}}x_\mu)^{N-1-j}\tilde M_j\,.
\ee
Following now procedure in \cite{SergyeyevKrtous:2008}, and using the trivial fact that
$[\tilde M_\mu, (-\iota_{\lst{\nu}}x_\nu)^{N-1-j}]=0$, we establish that
\ba
&&\hspace{-0.5cm}\sum_{j,k=0}^{N-1}(-i)^{j+k}(-\iota_{\lst{\mu}}x_\mu)^{N-1-j}(-\iota_{\lst{\nu}}x_\nu)^{N-1-k}\times\nonumber\\
&&\hspace{4.5cm}\times[\tilde M_j,\tilde M_k]=0\,,
\ea
from which Eq.~\eqref{com} follows.

\subsection{$R$-representation and standard separability}
We have seen that the new operators $\tilde M_j$ \eqref{RMR} possess a remarkable property---they can be expressed in the form \eqref{tildeMj}, where the
operators $\tilde M_\mu$ act only on the spinor $\chi_\mu$. Hence, such operators are directly related to standard tensorial separability. Indeed,
a solution of
\be\label{eigenvaluetilde}
K_k \psi = i\,\psc{k}\psi\;,\quad \tilde M_j \psi = \chc{j}\psi\;,
\ee
can be found in the standard tensorial separated form
\begin{equation}\label{tens_sep}
    \psi = \exp\bigl({\textstyle i\sum_k\psc{k}\psi_{k}}\bigr)\,
           \bigotimes_\nu \chi_\nu\;,
\end{equation}
where $\chi_\nu$ satisfy the equation \eqref{chieq}. This can be easily seen by calculating $\tilde M_\mu \psi$ while using Eq. \eqref{tildeMmu}. Note also
that the R-separability discussed in Sec.~\ref{sc:sep} is recovered by applying $R$ on the l.h.s. of \eqref{eigenvaluetilde}.

Moreover, operators $\tilde M_j$ are nothing else but operators \eqref{opsM} in the `R-representation'
in which we take
\be
\tilde \gamma^a=R^{-1}\gamma^a R\,,
\ee
with $\gamma^a$ defined earlier.

\section{Conclusions}\label{concl}
The Dirac equation in Kerr-NUT-(A)dS spacetimes in all dimensions possesses a truly remarkable property. Namely, its solution can be found by separating variables
and the resulting ordinary differential equations can be completely decoupled.

We have demonstrated that behind the separability stands a complete set of first-order mutually commuting operators
that can be generated from the PCKY tensor, present in the spacetime geometry.  These results directly generalize the corresponding
results on separability of the Hamilton--Jacobi and Klein--Gordon equations and further establish the
unique role which the PCKY tensor plays in determining the remarkable properties of the Kerr-NUT-(A)dS geometry in all dimensions.

A very important open question left for the future is whether the PCKY tensor is also intrinsically linked to other higher-spin perturbations.
In particular, can the electromagnetic and gravitational perturbations in general rotating higher-dimensional Kerr-NUT-(A)dS spacetimes
be decoupled and separated?

\section*{Acknowledgments}
We are grateful to G.W. Gibbons and C.M. Warnick for discussions and
reading the manuscript. P.K. was supported by by Grant No.~GA\v{C}R-202/08/0187 and
Project No.~LC06014 of the Center of Theoretical Astrophysics.
D.K. is the Clare College Research Associate and is grateful to the Herchel Smith Postdoctoral Fellowship at the University of Cambridge for financial support.

\appendix

\section{}\label{apx}

\subsection{Spin connection}\label{apx:spincon}
In even dimension the only non-zero connection (Ricci) coefficients with respect to the frame ${\ef^\mu,\,\ef^{\hat\mu}}$ are:\pagebreak[0]
\begin{equation}\label{conncoef}
\begin{split}
  &\omega_{\mu\mu\nu} = -\omega_{\mu\nu\mu}
    =\sqrt{\frac{X_\nu}{U_\nu}}\,\frac{x_\nu}{x_\nu^2-x_\mu^2}\;,\\
  &\omega_{\mu\hat\mu\hat\nu} = -\omega_{\mu\hat\nu\hat\mu}
    =\sqrt{\frac{X_\nu}{U_\nu}}\,\frac{x_\mu}{x_\nu^2-x_\mu^2}\;,\\
  &\omega_{\hat\mu\hat\mu\nu} = -\omega_{\hat\mu\nu\hat\mu}
    =\sqrt{\frac{X_\nu}{U_\nu}}\,\frac{x_\nu}{x_\nu^2-x_\mu^2}\;,\\
  &\omega_{\hat\mu\hat\nu\mu} = -\omega_{\hat\mu\mu\hat\nu}
    =\sqrt{\frac{X_\nu}{U_\nu}}\,\frac{x_\mu}{x_\nu^2-x_\mu^2}\;,\\
  &\omega_{\hat\mu\nu\hat\nu} = -\omega_{\hat\mu\hat\nu\nu}
    =\sqrt{\frac{X_\mu}{U_\mu}}\,\frac{x_\nu}{x_\nu^2-x_\mu^2}\;,\\
  &\omega_{\hat\mu\hat\mu\mu} = -\omega_{\hat\mu\mu\hat\mu}
    =\frac12\sqrt{\frac{X_\mu}{U_\mu}}\frac{X_\mu'}{X_\mu}+
           \sqrt{\frac{X_\mu}{U_\mu}}\sum_{\substack{\nu\\\nu\neq\mu}}\frac{x_\mu}{x_\nu^2-x_\mu^2}\;.
\end{split}\raisetag{25ex}
\end{equation}
Here, indices ${\mu}$ and ${\nu}$ are different. In odd dimension the same Ricci coefficients apply, plus the following extra terms:
\begin{equation}\label{conncoef_odd}
\begin{split}
  &\omega_{\mu \hat\mu 0} = -\omega_{\mu 0 \hat\mu} = - \frac{\sqrt{S}}{x_\mu}  \; , \\
  &\omega_{\hat\mu \mu 0} = -\omega_{\hat\mu 0 \mu} = \frac{\sqrt{S}}{x_\mu} \; , \\
  &\omega_{0 \mu 0} = -\omega_{0 0 \mu} = - \sqrt{\frac{X_\mu}{U_\mu}} \frac{1}{x_\mu} \; , \\
  &\omega_{0\hat\mu\mu} = -\omega_{0\mu\hat\mu} = - \frac{\sqrt{S}}{x_\mu} \; . \\
\end{split}\raisetag{25ex}
\end{equation}

\subsection{Useful Identities}\label{apx:ident}
We generalize the definition of functions $\Bo{k}_\mu$ and $\Bo{k}$, \eqref{Bdef}, as following:
\be
\Bo{k}_{\mu_1\dots\mu_j} = \sum\limits_{\substack{\nu_1,\dots,\nu_k\\\nu_1<\dots<\nu_k \\ \nu_i\ne \mu_1, \dots , \mu_j  }}\!\!\!\!\!   \iota_{\lst{\nu_1}}x_{\nu_1}\cdots\ \iota_{\lst{\nu_k}}x_{\nu_k} \;.\label{generalised_Bdef}\\
\ee
Such functions obey
\be
\Bo{k}_{\mu_1 \dots \mu_j} = \Bo{k}_{\mu_1 \dots \mu_j \nu} + \iota_{\lst{\nu}} x_\nu \Bo{k-1}_{\mu_1 \dots \mu_j \nu} \, .  \label{eq:generalisedB_additivity}
\ee
Therefore we can write
\ba
&& \hspace{-1cm}\sum_{\substack{\nu\\\nu \neq \mu_1, \dots , \mu_j }} \Bo{k}_{\mu_1 \dots \mu_j\nu} = \nn \\
&& =\sum_{\substack{\nu\\\nu \neq \mu_1, \dots , \mu_j }}  \left( \Bo{k}_{\mu_1 \dots \mu_j} - \iota_{\lst{\nu}} x_\nu \Bo{k-1}_{\mu_1 \dots \mu_j\nu} \right)  \nn \\
&&  =(N-j-k) \Bo{k}_{\mu_1 \dots \mu_j} \, . \label{eq:generalisedB_sum}
\ea
As a direct consequence of Eqs.~\eqref{eq:generalisedB_additivity} and \eqref{eq:generalisedB_sum} we can derive the following important
relations used in the main text:
\ba
&&\sum_\mu \Bo{j-1}_\mu=(N-j+1)\Bo{j-1}\,,\label{dveid}\\
&&\sum_{\substack{\nu\\\nu\neq\mu}}\! \frac{x_\mu\!+\!\iota_{\lst{\mu\nu}}x_\nu}{x_\mu^2\!-\!x_\nu^2}
         \bigl(\Bo{j}_\nu\!-\!\Bo{j}_\mu\bigl)
         = (N\!-\!j)\,\iota_{\lst{\mu}}\,\Bo{j{-}1}_\mu,\label{eq:Bid}\ \,\,\\
&&\sum_{j=0}^N \Bo{j} \left( -\iota_{\lst{\mu}} x_\mu \right)^{N-1-j} = 0 \, ,\label{eq:sum_Boj}\\
&&\sum_{j=0}^N \Bo{j-1} \left( -\iota_{\lst{\mu}} x_\mu \right)^{N-1-j} = - \frac{\Bo{N}}{x_\mu^2} \, . \label{eq:sum_Boj-1}
\ea
Another important relation is
\be
 \sum_\mu \frac{\Bo{j}_\mu}{{\iota_{\lst{\mu}}x_\mu} \Vo_{\mu}}=\frac{{\Bo{j}}}{\Bo{N}}\,,
 \ee
which, together with \eqref{eq:generalisedB_additivity} and the equality
\be   \label{sum_1over_x_squared_Vo}
\sum_\mu \frac{1}{x_\mu^2 \Vo_\mu}=\sum_\mu \frac{1}{(\iota_{\lst{\mu}}x_\mu)^2 \Vo_\mu} =  \frac{1}{\Bo{N}} \sum_\mu \frac{1}{\iota_{\lst{\mu}} x_\mu} \,
\ee
mentioned in \cite{OotaYasui:2008}, can be used to prove that $\Bo{j-1}$ can be expressed as
\be\label{id15}
\Bo{j-1}=\Bo{j}\sum_\mu \frac{1}{\iota_{\lst{\mu}}x_\mu}-\Bo{N}\sum_\mu\frac{\Bo{j}_\mu}{\Vo_{\mu}x_\mu^2}\,.
\ee

Let us finally state two important relations including the $R$ factor. Using the fact that
$U_\mu=(\Vo_\mu\sigma_{\lst{\mu}})^2$, we can derive that
\be\label{eq:gamma_mu_phi}
R^{-1}\gamma^\mu R =  \frac{\sqrt{|U_\mu|}}{\Vo_{\mu}} \left( - \iota_{\lst{\mu}}\right)^{N-\mu} \sigma_{\lst{\mu}} \, ,
\ee
which is an operator analogue of Eq.~(20) in \cite{OotaYasui:2008}. For the derivative of the factor $R$ one gets
\ba\label{derPhi}
 &&  \Bigl(\cd{x_\mu} \gamma^\mu R\Bigr)= \gamma^\mu \cd{x_\mu} R = \nn \\
&&  =\gamma^\mu \cd{x_\mu} \biggl(  \prod_{\substack{\nu\\\nu<\mu}}\bigl(x_\nu{+}\iota_{\lst{\mu\nu}}x_\mu\bigr)^{\!-\frac12}
     \prod_{\substack{\nu\\\mu<\nu}}\bigl(x_\mu{+}\iota_{\lst{\mu\nu}}x_\nu\bigr)^{\!-\frac12} \times \nn \\
 && \hspace{1.8cm} \prod_{\substack{\kappa,\lambda\\\kappa<\lambda;\;\kappa,\lambda\neq\mu}}
         \!\!\bigl(x_\kappa{+}\iota_{\lst{\kappa\lambda}}x_\lambda\bigr)^{\!-\frac12} \biggr)  \nn \\
 &&= \Bigl( - \frac12 \sum_{\substack{\nu\\\nu\neq\mu}} \frac1{x_\mu\!{-}\iota_{\lst{\mu\nu}}x_\nu} \Bigr) \gamma^\mu R \, .
\ea



\newpage
\providecommand{\href}[2]{#2}\begingroup\raggedright\endgroup

\end{document}